\documentstyle[seceq,preprint]{jpsj}

\newcommand{\vecvar}[1]{\mbox{\boldmath$#1$}}
\newcommand{\lsim}{ < \kern -11.8pt \lower 5pt \hbox{$\displaystyle \sim$}}
\newcommand{\gsim}{ > \kern -12pt   \lower 5pt   \hbox{$\displaystyle \sim$}}

\def\runtitle{Orbital Magnetism and Current Distribution 
of Two-Dimensional Electrons under Confining Potential}

\def\runauthor{Yasushi {\sc Ishikawa} and Hidetoshi {\sc Fukuyama}}

\title
{Orbital Magnetism and Current Distribution \\
 of Two-Dimensional Electrons under Confining Potential}
\author
{ 
Yasushi {\sc Ishikawa}\footnote{E-mail: ishikawa@watson.phys.s.u-tokyo.ac.jp}
and Hidetoshi {\sc Fukuyama}
}

\inst
{
Department of Physics, University of Tokyo, Tokyo 113-0033
}

\recdate
{\today}

\abst
{
The spatial distribution of electric current under magnetic field
and the resultant orbital magnetism
have been 
studied for two-dimensional electrons under a harmonic confining potential
$\:V(\vecvar{r})=m \omega_0^2 r^2/2\:$
in various regimes of temperature and magnetic field, and
the microscopic conditions for the validity of Landau diamagnetism are
clarified. 
Under a weak magnetic field 
$(\:\omega_c\lsim\:\:\omega_0$, $\omega_c$
 being a cyclotron frequency )
and at low temperature $(\:T\lsim\:\:\hbar\omega_0\:)$, 
where the orbital magnetic moment fluctuates as a function of
the field, the currents are irregularly distributed paramagnetically or 
diamagnetically inside the bulk region.
As the temperature is raised under such a weak field, 
however, the currents in the bulk region are
immediately reduced and finally 
there only remains the diamagnetic current flowing 
along the edge. At the same time, 
the usual Landau diamagnetism results for the total
magnetic moment.
The origin of this dramatic temperature dependence is seen to be in 
the multiple reflection of electron waves by the boundary confining potential,
which becomes important once
the coherence length of electrons gets longer
than the system length.
Under a stronger field $(\:\omega_c\gsim\:\:\omega_0\:)$, on the other hand,
the currents in the bulk region cause 
de Haas-van Alphen effect at low temperature as $T\lsim\:\:\hbar\omega_c$.
As the temperature gets higher $(\:T\gsim\:\:\hbar\omega_c\:)$ 
under such a strong field, 
the bulk currents are reduced and the Landau diamagnetism by the edge current
is recovered.
}

\kword
{orbital magnetism, confining potential, Landau diamagnetism, 
edge current, multiple reflection, mesoscopic system}

\begin{document}
\newread\epsffilein    
\newif\ifepsffileok    
\newif\ifepsfbbfound   
\newif\ifepsfverbose   
\newif\ifepsfdraft     
\newdimen\epsfxsize    
\newdimen\epsfysize    
\newdimen\epsftsize    
\newdimen\epsfrsize    
\newdimen\epsftmp      
\newdimen\pspoints     
\pspoints=1bp          
\epsfxsize=0pt         
\epsfysize=0pt         
\def\epsfbox#1{\global\def\epsfllx{72}\global\def\epsflly{72}%
   \global\def\epsfurx{540}\global\def\epsfury{720}%
   \def\lbracket{[}\def\testit{#1}\ifx\testit\lbracket
   \let\next=\epsfgetlitbb\else\let\next=\epsfnormal\fi\next{#1}}%
\def\epsfgetlitbb#1#2 #3 #4 #5]#6{\epsfgrab #2 #3 #4 #5 .\\%
   \epsfsetgraph{#6}}%
\def\epsfnormal#1{\epsfgetbb{#1}\epsfsetgraph{#1}}%
\def\epsfgetbb#1{%
%
%
\openin\epsffilein=#1
\ifeof\epsffilein\errmessage{I couldn't open #1, will ignore it}\else
%
%
   {\epsffileoktrue \chardef\other=12
    \def\do##1{\catcode`##1=\other}\dospecials \catcode`\ =10
    \loop
       \read\epsffilein to \epsffileline
       \ifeof\epsffilein\epsffileokfalse\else
%
%
          \expandafter\epsfaux\epsffileline:. \\%
       \fi
   \ifepsffileok\repeat
   \ifepsfbbfound\else
    \ifepsfverbose\message{No bounding box comment in #1; using defaults}\fi\fi
   }\closein\epsffilein\fi}%
%
%
\def\epsfclipon{\def\epsfclipstring{ clip}}%
\def\epsfclipoff{\def\epsfclipstring{\ifepsfdraft\space clip\fi}}%
\epsfclipoff
\def\epsfsetgraph#1{%
   \epsfrsize=\epsfury\pspoints
   \advance\epsfrsize by-\epsflly\pspoints
   \epsftsize=\epsfurx\pspoints
   \advance\epsftsize by-\epsfllx\pspoints
%
%
   \epsfxsize\epsfsize\epsftsize\epsfrsize
   \ifnum\epsfxsize=0 \ifnum\epsfysize=0
      \epsfxsize=\epsftsize \epsfysize=\epsfrsize
      \epsfrsize=0pt
%
%
     \else\epsftmp=\epsftsize \divide\epsftmp\epsfrsize
       \epsfxsize=\epsfysize \multiply\epsfxsize\epsftmp
       \multiply\epsftmp\epsfrsize \advance\epsftsize-\epsftmp
       \epsftmp=\epsfysize
       \loop \advance\epsftsize\epsftsize \divide\epsftmp 2
       \ifnum\epsftmp>0
          \ifnum\epsftsize<\epsfrsize\else
             \advance\epsftsize-\epsfrsize \advance\epsfxsize\epsftmp \fi
       \repeat
       \epsfrsize=0pt
     \fi
   \else \ifnum\epsfysize=0
     \epsftmp=\epsfrsize \divide\epsftmp\epsftsize
     \epsfysize=\epsfxsize \multiply\epsfysize\epsftmp   
     \multiply\epsftmp\epsftsize \advance\epsfrsize-\epsftmp
     \epsftmp=\epsfxsize
     \loop \advance\epsfrsize\epsfrsize \divide\epsftmp 2
     \ifnum\epsftmp>0
        \ifnum\epsfrsize<\epsftsize\else
           \advance\epsfrsize-\epsftsize \advance\epsfysize\epsftmp \fi
     \repeat
     \epsfrsize=0pt
    \else
     \epsfrsize=\epsfysize
    \fi
   \fi
%
%
   \ifepsfverbose\message{#1: width=\the\epsfxsize, height=\the\epsfysize}\fi
   \epsftmp=10\epsfxsize \divide\epsftmp\pspoints
   \vbox to\epsfysize{\vfil\hbox to\epsfxsize{%
      \ifnum\epsfrsize=0\relax
        \includegraphics{\ifepsfdraft}%
      \else
        \epsfrsize=10\epsfysize \divide\epsfrsize\pspoints
        \includegraphics{\ifepsfdraft}%
      \fi
      \hfil}}%
\global\epsfxsize=0pt\global\epsfysize=0pt}%
%
%
{\catcode`\%=12 \global\let\epsfpercent=
%
%
\long\def\epsfaux#1#2:#3\\{\ifx#1\epsfpercent
   \def\testit{#2}\ifx\testit\epsfbblit
      \epsfgrab #3 . . . \\%
      \epsffileokfalse
      \global\epsfbbfoundtrue
   \fi\else\ifx#1\par\else\epsffileokfalse\fi\fi}%
%
%
\def\epsfempty{}%
\def\epsfgrab #1 #2 #3 #4 #5\\{%
\global\def\epsfllx{#1}\ifx\epsfllx\epsfempty
      \epsfgrab #2 #3 #4 #5 .\\\else
   \global\def\epsflly{#2}%
   \global\def\epsfurx{#3}\global\def\epsfury{#4}\fi}%
%
%
\def\epsfsize#1#2{\epsfxsize}
%
%
\let\epsffile=\epsfbox

\sloppy
\maketitle

\section{Introduction}
\label{L1}
Bohr-van Leeuwen's theorem tells us that
the orbital magnetism does not appear 
in classical theory~\cite{Bohr,vanVleck}.
The physical argument for this fact is that 
the diamagnetic current due to cyclotron orbits of
electrons in the bulk region
is perfectly cancelled by the paramagnetic current due to skipping orbits 
near the boundary.  
The orbital magnetism,
which is then possible only in quantum mechanics, 
was originally derived based on the Landau
levels~\cite{Landau1}.
In the actual derivation, 
effects of the boundary
have not been taken into account explicitly.
These effects of boundary on the orbital magnetism 
have been later investigated by many authors
~\cite{Kubo,DY5,DY6,DY7,DY8,DY9,Robnik,DY12,DY13,NA,DY,Shapiro}.
Above all, Kubo~\cite{Kubo} has applied the Wigner representation to an electron system
under a magnetic field and
shown that, if a confining potential is slowly varying in space 
compared to the electron wave length, the Landau diamagnetism results. 
However, this treatment is not valid at low temperature
as stressed by Kubo and
clear from the expansion parameters.
The magnetic moment at $\:T=0\:$, on the other hand, has been shown 
by Denton~\cite{DY7} and Nemeth~\cite{DY12} to be different from the Landau
diamagnetism in a system under a harmonic confining potential. 
Yoshioka and Fukuyama~\cite{DY} has actually indicated that under a weak field and
at low temperature $(\:k_B T\lsim\:$ energy spacing by a confinement),
the magnetic moment of the whole system shows a large fluctuation, and 
as the temperature is raised under such a weak field, fluctuations
disappear and Landau diamagnetism is recovered.
Hajdu and Shapiro~\cite{Shapiro} pointed out 
by studying the case of a groove with a width $L$
that the temperature below which such fluctuations appear
is such that $k_B T\sim \hbar/\tau_{tr}$ where $\tau_{tr}=L/v_F$, 
the time of flight for
electrons at the Fermi energy across the system.  
Stimulated by these indications, 
we will study in this paper the spatial distribution of current 
in a confined system in order to understand more details of
such a variety of orbital magnetism.
From the investigations so far,
the shape of confining potential, 
whether it is harmonic or hard wall,
is expected not to affect
qualitative aspects of the characteristic features of the orbital magnetism.
Therefore we assume a harmonic potential 
which makes the analytical studies possible,
and then clarify the relationship 
between the orbital magnetism and the spatial distribution of current
in various regimes of temperature and magnetic field.

The organization of this paper is as follows.
In \S\ref{L2}, 
we introduce the model and summarize the general feature of the orbital magnetic
moment.
In \S\ref{L3}, the spatial distribution of current will be studied
in various regions of temperature and magnetic field,
and the microscopic conditions for the validity of Landau diamagnetism  
are clarified.
Summary is given in \S\ref{L4}.

We take units $k_B=1$ in the following.
\section{Magnetic Moment of Two-Dimensional Confined System}
\label{L2}
In this section, we introduce our model and 
explain the general properties of the magnetic moment of the system. 
We consider two-dimensional electrons confined by an isotropic harmonic potential
and the magnetic field is applied perpendicular to the system.
For simplicity, we consider spinless electrons with the total number $N_0$
and neglect Coulomb interactions 
between them.
We assume $N_0$ is large enough that the difference between
a grand canonical ensemble and a canonical one is of no importance
and rely on a grand canonical one to derive a magnetic moment. 
The word ``magnetic moment'' in this paper implies a magnetic moment of 
the whole system.

The Hamiltonian is written as
\begin{equation}
{\cal H} =\frac{1}{2m}\left(\vecvar{p}+\frac{e}{c}\vecvar{A}\right)^2+
\frac{1}{2}\makebox[0.1em]{}m\makebox[0.1em]{}\omega_0^2\makebox[0.1em]{}\vecvar{r}^2,
\end{equation}
where $\vecvar{p}$ and $\vecvar{r}$ are two-dimensional vector,
$m$ is the electron mass and $(-e)$ is the electron charge.
The radius of the system, $R$, is classically defined as
\begin{equation}
\frac{1}{2}\makebox[0.1em]{}m\makebox[0.1em]{}
\omega_0^2\makebox[0.1em]{} R^2=\mu,
\label{Ru}\end{equation}
where $\mu$ is the chemical potential.

By taking a symmetric gauge $\vecvar{A}=\frac{1}{2}\vecvar{H}\times\vecvar{r}$,  
we can obtain an eigenfunction diagonal with respect to the angular momentum 
$\alpha$ as
\begin{eqnarray}
\psi_{n\alpha}(\vecvar{r})&=&
\frac{{\rm e}^{{\rm i}\alpha\theta}}{\sqrt{2\pi}}
\:R_{n \alpha}(r)\nonumber\\
R_{n \alpha}(r)&=&\frac{1}{l}\sqrt{\frac{n\makebox[0.2em]{}!}
{(\makebox[0.2em]{}n+|\alpha|\makebox[0.2em]{})\makebox[0.2em]{}!}}
\:\exp\left[-\frac{r^2}{4\makebox[0.1em]{}l^2}\right]
\left(\frac{r}{\sqrt{2}\makebox[0.1em]{}l}\right)^{|\alpha|}
L_n^{(|\alpha|)}\left[\frac{r^2}{2\makebox[0.1em]{}l^2}\right]
\label{WF}
\end{eqnarray}
where polar coordinates $(\:r,\:\theta\:)$ are used, and
$\:n=0,1,2,...$ , $\:\alpha=0, \pm 1, \pm 2,...$
and $l=\sqrt{\hbar/m\omega}
\makebox[0.3em]{},\makebox[0.3em]{}
\omega=\sqrt{\omega_c^2+(2\makebox[0.1em]{}\omega_0)^2}\makebox[0.3em]{},
\makebox[0.3em]{}
\omega_c=eH/mc\:$ being the cyclotron frequency and
$L_n^{(\alpha)}$ is the Laguerre polynomial.\\
The eigenenergy of this state, $E_{n \alpha}$, is given by
\begin{equation}
E_{n\alpha}=\hbar\omega_c\:\frac{\alpha}{2}+\hbar\omega
\left(n+\frac{|\alpha|+1}{2}\right).
\label{energy}
\end{equation}
Especially under a extremely strong field $(\omega_c\gg\omega_0)$,
this eigenenergy $E_{n\alpha}$ becomes
$\hbar\omega_c(n+1/2)$ for negative $\alpha$.
So, the quantum number $n$ corresponds to the Landau level index.  

By use of eq.~(\ref{energy}), the thermodynamic potential $\Omega$ is written as
\begin{equation}
\Omega=-\frac{1}{\beta}\sum_{n=0}^{\infty}\sum_{\alpha=-\infty}^{\infty}
\log\left[\makebox[0.2em]{}1+{\rm e}^{-\beta(E_{n \alpha}-\mu)}\right],
\label{Omega}
\end{equation}
where $\beta=1/T$ and
$\mu$ is the chemical potential adjusted to fix the average  
electron number to $N_0$ at each values of $H$ and $T$.\\
From the thermodynamic potential, the magnetic moment $M$ is given as
\begin{eqnarray}
M&=&-\left(\frac{\partial\Omega}{\partial H}\right)_{\mu}\nonumber\\
&=&\sum_{n \alpha}\left(-\frac{\partial E_{n\alpha}}{\partial H}\right)f(E_{n\alpha})
\label{Statistic}
\end{eqnarray}
where $f(E)$ is the Fermi distribution function.

Applying the Poisson summation formula~\cite{Landau2}
to the sum over $n$ and $\alpha$ in eq.~(\ref{Omega})
(for details, see Appendix~\ref{LA}),
one obtains 
\begin{equation}
\Omega=\Omega_0+\Omega_L+\Omega_{osc}\makebox[0.5em]{},
\label{EachOmega}
\end{equation}
where
\begin{eqnarray}
\Omega_0&=&-\frac{1}{\beta(\hbar\omega_0)^2}
\int_{0}^{\infty}{\rm d}\varepsilon
\int_{0}^{\infty}{\rm d}\eta\makebox[0.2em]{}
\log\left[\makebox[0.2em]{}1+{\rm e}^{-\beta(\varepsilon+\eta-\mu)}\right]
+\frac{1}{12}\makebox[0.2em]{}\mu\makebox[0.5em]{},
\label{Omega0}\\[1.5em]
\Omega_L&=&\frac{1}{24}\left(\frac{\omega_c}{\omega_0}\right)^2\mu
\makebox[0.5em]{},\\[1.5em]
\Omega_{osc}&\simeq&\frac{1}{2\pi\beta}\sum_{k=1}^{\infty}(-1)^k\left[
\left(\frac{\omega}{\omega_0}\right)^2\frac{1}{k^2}-\frac{\pi^2}{3}\right]
\frac{\sin\left(2\pi k\mu/\hbar\omega\right)}
{\sinh\left(2\pi^2k/\beta\hbar\omega\right)}\nonumber\\
&+&\frac{1}{2\pi\beta}\sum_{\sigma=\pm}\sum_{l=1}^{\infty}
\left(\frac{\omega_{\sigma}}{\omega}\right)\frac{1}{l^2}
\frac{\sin\left(2\pi l\mu/\hbar\omega_{\sigma}\right)}
{\sinh\left(2\pi^2l/\beta\hbar\omega_{\sigma}\right)}\nonumber\\
&+&\frac{1}{\pi\beta}\sum_{\sigma=\pm}\sum_{k=1}^{\infty}\sum_{l=1}^{\infty}
\frac{(-1)^k}{l}
\left[\makebox[0.5em]{}
\frac{\sin\left[\frac{\pi\mu}{\hbar\omega}\left(k-\frac{\omega}{\omega_{\sigma}}l
\right)\right]
\cos\left[\frac{\pi\mu}{\hbar\omega}\left(k+\frac{\omega}{\omega_{\sigma}} l\right)\right]}
{\left(k-\frac{\omega}{\omega_{\sigma}} l\right)
\sinh(2\pi^2l/\beta\hbar\omega_{\sigma})}\right.\nonumber \\
&&\makebox[9.2em]{}+
\left.\frac{\sin\left[\frac{\pi\mu}{\hbar\omega}\left(k+\frac{\omega}{\omega_{\sigma}}l
\right)\right]
\cos\left[\frac{\pi\mu}{\hbar\omega}\left(k-\frac{\omega}{\omega_{\sigma}} l\right)\right]}
{\left(k+\frac{\omega}{\omega_{\sigma}} l\right)
\sinh(2\pi^2l/\beta\hbar\omega_{\sigma})}\makebox[0.5em]{}\right],
\label{osc}
\end{eqnarray}
where $\omega_{\pm}=(\omega\pm\omega_c)/2$ and the relation $\mu\gg T$ is assumed. 
In the derivation of oscillatory terms in $\Omega_{osc}$, 
we used the approximation as follows,
\begin{eqnarray} 
&&\int_{-\beta(\mu-\eta)}^{\infty}{\rm d}\xi\makebox[0.2em]{}
\frac{{\rm e}^{\xi}}{({\rm e}^{\xi}+1)^2}\cdot
{\rm e}^{{\rm i}\frac{2\pi k}{\beta \hbar\omega}\xi}\nonumber\\
&\simeq&
\int_{-\infty}^{\infty}{\rm d}\xi\makebox[0.2em]{}
\frac{{\rm e}^{\xi}}{({\rm e}^{\xi}+1)^2}\cdot
{\rm e}^{{\rm i}\frac{2\pi k}{\beta \hbar\omega}\xi}
\cdot \theta\makebox[0.1em]{}[\makebox[0.1em]{}\mu-\eta\makebox[0.1em]{}],
\end{eqnarray}
where $ \theta\makebox[0.1em]{}[\makebox[0.1em]{}x\makebox[0.1em]{}]$ is 
the Heaviside function.

As is known from eq.~(\ref{Omega0}),
$\Omega_0$ is dependent on a magnetic field only through the chemical potential 
$\mu(H,\makebox[0.2em]{}T)$, which is regarded as a function of the field
and temperature and almost a constant as a function of the field 
when $\mu/\hbar\omega_c\gg 1$.
Therefore the contribution of $\Omega_0$ to the magnetic moment is much smaller
than those due to $\Omega_L$ and $\Omega_{osc}$.

Based on this result, the field dependence of the magnetic moment 
is classified into three regions;
`` Mesoscopic Fluctuation (MF)'',`` Landau Diamagnetism (LD) '' 
and  `` de Haas-van Alphen (dHvA)'', as is shown in Fig.~\ref{Phase}.
``MF'' corresponds to the region as 
$\:T\lsim\makebox[0.3em]{}\hbar\omega_{-}\:$, which
implies $\:T/\hbar\omega_0\lsim\makebox[0.3em]{}1\:$ under a weak field 
$(\:\omega_c/\omega_0\lsim\makebox[0.3em]{}1\:)$ and 
$\makebox[0.3em]{}T/\hbar\omega_0\lsim\makebox[0.3em]{}
(\omega_c/\omega_0)^{-1}$ 
under a strong field
$(\:\omega_c/\omega_0\gsim\makebox[0.3em]{}1\:)$, while
``LD'' corresponds to the region as $T\gsim\makebox[0.3em]{}\hbar\omega_{+}$, 
which requires $\:T/\hbar\omega_0\gsim\makebox[0.3em]{}1\:$ under a weak field 
$(\:\omega_c/\omega_0\lsim\makebox[0.3em]{}1\:)$ and 
$\:T/\hbar\omega_0\gsim\makebox[0.3em]{}\omega_c/\omega_0\:$ 
under a strong field
$(\:\omega_c/\omega_0\gsim\makebox[0.3em]{}1\:)$.
The other region in Fig.~\ref{Phase} is ``dHvA''.

Fig.~\ref{Moment} shows field dependences of magnetic moment
at various temperatures; 
(a),(b) and (c),(d) are respectively under a weak field 
$(\:\omega_c\lsim\:\:\omega_0\:)$ 
and a strong field $(\:\omega_c\gsim\:\:\omega_0\:)$.
$N_0$ is set at $5000$ in this calculation.
At first, we focus on the region of a weak field 
$(\:\omega_c\lsim\:\:\omega_0\:)$ shown in (a) and (b), 
where (a) corresponds to ``MF'' and (b) ranges from ``MF'' to ``LD''. 
At low temperature as $T\lsim\:\:\hbar\omega_0$ corresponding to ``MF'',
the magnetic moment shows a large fluctuation with respect to the field,
as Yoshioka and Fukuyama have shown.
In this ``MF'' region, all the oscillatory terms in 
$\Omega_{osc}$ contribute to the magnetic moment leading to 
such a large fluctuation, 
in addition to $\Omega_{L}=-1/2\cdot \chi_{L}H^2$ 
$(\: \chi_{L}=-1/3\cdot D_0 \mu_{B}^2$, the Landau diamagnetic susceptibility, 
where $D_0=\mu/(\hbar\omega_0)^2$, 
the density of states at Fermi energy ), which leads to the usual Landau diamagnetism.
Particularly at much lower temperature $(\:T\ll\makebox[0.3em]{}\hbar\omega_0\:)$
and  under a much weaker field $(\:\omega_c\ll\omega_0\:)$,
the magnetic moment shows a strong paramagnetism, which was first
noticed by Meier and Wyder~\cite{DY8} and discussed by Budzin et al.~\cite{DY9}
This strong paramagnetism is attributed to the rotational symmetry of the system.
Hence, in the presence of weak disorder,
a large spatial variation of magnetic moment, either paramagnetic or diamagnetic,
is expected, which fact is the cause of a large variance of orbital susceptibility
in the limit of weak magnetic field at low temperature
\cite{DY17,HF,DY18,DY19,HY}.
As the temperature is raised $(\:T\gsim\:\:\hbar\omega_0\:)$ under such a weak field
$(\:\omega_c\lsim\:\:\omega_0\:)$,
the fluctuation of magnetic moment is reduced and the magnetic moment becomes 
linearly dependent on the field, the slope of which gives the Landau diamagnetic 
susceptibilities $\chi_{L}$ corresponding to ``LD''.
In this ``LD'' region, the contribution of $\Omega_{osc}$ are reduced
and $\Omega_L$ becomes dominant.

Fig.~\ref{Moment}(c) and (d) show the field dependence of magnetic moment 
under a strong field $(\:\omega_c\gsim\:\:\omega_0\:)$ at various temperatures.
(c) and (d) range from ``MF'' to ``dHvA'' and from ``dHvA'' to ``LD'', respectively.
At low temperature as $T\lsim\:\:\hbar\omega_{-}\simeq \hbar\omega_0^2/\omega_c$, 
magnetic moment shows slow oscillation with a large amplitude
and rapid oscillation with a small amplitude, as is shown in Fig.~\ref{Moment}(c).
The former oscillation with respect to the field is characterized by 
$\mu/\hbar\omega\simeq\mu/\hbar\omega_c$, 
which is caused by the periodic intersections of chemical potential $\mu$
by the Landau level ( the states characterized by $n$ in eq.~(\ref{energy}) )
and corresponds to the de Haas-van Alphen oscillation.
The latter oscillation with respect to the field is governed by 
$\phi/\phi_0$ where $\phi$ is the total flux penetrating the system
and $\phi_0=hc/e$, flux quantum.
When the total magnetic flux is increased by $\phi_0$,
the degeneracy of each Landau level under Fermi energy 
is increased by unity.
This causes the oscillation of the magnetic moment with a period $\phi_0$
as a function of $\phi$ 
at such a low temperature as 
$\:T\lsim\:\:\hbar\omega_{-}\simeq \hbar\omega_0^2/\omega_c\:$
which is energy spacing between different angular momentum states at each Landau level.
Physically, this oscillation is caused by a coherent motion of electrons 
along the edge and can be considered as the Aharonov-Bohm oscillation. 
Such a behavior under a strong field
$(\:\omega_c\gsim\makebox[0.3em]{}\omega_0\:)$ has been noted at $T=0$ 
by Meir, Wohlman, and Gefen~\cite{NA}. 
As the temperature is raised so as 
$\:\hbar\omega_-\lsim\:\:T\lsim\:\:\hbar\omega_+\simeq\hbar\omega_c\:$ 
under such a strong field $(\:\omega_c\gsim\:\:\omega_0\:)$ corresponding
to ``dHvA'', the AB oscillation disappears and there only remains 
the dHvA oscillation. 
In this ``dHvA'' region, the first term in $\Omega_{osc}$ 
has an appreciable contribution
in addition to $\Omega_L$.
At much higher temperature $(\:T\gsim\:\:\hbar\omega_+\simeq\hbar\omega_c\:)$, 
the dHvA oscillation disappears and 
magnetic moment shows a linear field dependence, i.e. the Landau diamagnetism
as seen in Fig.~\ref{Moment}(d) e.g. $T/\hbar\omega_0=4$. 



\section{Spatial Distribution of Current and Magnetic Moment}
\label{L3}
In this section, we study the relationship between a spatial distribution of current 
and a magnetic moment of the whole system, the latter of which
is given by the thermodynamic potential $\Omega$
in the previous section. 

The spatial distribution of current in the system is given as follows,
\begin{eqnarray}
\vecvar{J}(\vecvar{r})&=&{\rm Re}
\left< 
\hat{\psi}^{\dag}(\vecvar{r})
\frac{(-e)}{m}\left(
\hat{\vecvar{p}}(\vecvar{r})
+\frac{e}{c}\vecvar{A}(\vecvar{r})\right)
\hat{\psi}(\vecvar{r})
\right> \nonumber\\[2mm]
&=&J_{\theta}(r)\makebox[0.2em]{}\vecvar{e}_{\theta},
\end{eqnarray}
where $<\cdot\cdot\cdot>$ denotes the thermal average and $\hat{\psi}(\vecvar{r})$ is the
field operator and $\vecvar{e}_{\theta}=\frac{\partial\vecvar{r}}{\partial\theta}/
|\frac{\partial\vecvar{r}}{\partial\theta}|$ as shown in Fig.~\ref{System}.
By use of the eigenstates in eq.~(\ref{WF}) as the basis, 
$J_{\theta}(r)$ is given as
\begin{equation}
J_{\theta}(r)=(-e)v_0\sum_{n\alpha}\left[
\alpha\left(\frac{r}{\xi}\right)^{-1}+\frac{\omega_c}
{2\makebox[0.1em]{}\omega_0}\left(\frac{r}{\xi}\right)
\right]R_{n\alpha}(r)^{2}\makebox[0.3em]{}f(E_{n\alpha}),
\label{Jr}
\end{equation} 
where $\xi=\sqrt{\hbar/m\makebox[0.1em]{}\omega_0}\makebox[0.3em]{}$, 
the characteristic length, and
$v_0=\omega_0\makebox[0.1em]{}\xi\makebox[0.3em]{}$, 
the characteristic velocity of electrons.\\
This local current $\vecvar{J}(\vecvar{r})$ induces 
a magnetic moment $\vecvar{M}(\vecvar{r})$ given as follows,
\begin{eqnarray}
\vecvar{M}(\vecvar{r})&=&\frac{1}{2c}\vecvar{r}\times\vecvar{J}(\vecvar{r})\nonumber\\
&=&\frac{1}{2c}\makebox[0.2em]{}r\makebox[0.2em]{}J_{\theta}(r)\makebox[0.2em]{}
\vecvar{e}_{z}\nonumber\\[2mm]
&\equiv& M_z(r)\makebox[0.2em]{}\vecvar{e}_{z}.
\end{eqnarray}
Therefore, by integrating this magnetic moment with respect to $\vecvar{r}$, 
we get the relation between the total magnetic moment $M$ ($z$-component) and
the local current $J_{\theta}(r)$ as 
\begin{eqnarray} 
M&=&\int{\rm d}S\makebox[0.3em]{} M_{z}(r)\nonumber\\
&=&\frac{\pi}{c}\int_{0}^{\infty}{\rm d}r\makebox[0.3em]{} r^2 J_{\theta}(r).
\label{MJ}
\end{eqnarray}
This expression of magnetic moment derived from the local current density 
coincides with the one in eq.~(\ref{Statistic})
derived from the thermodynamic potential (see Appendix~\ref{LB}).


\subsection{Region of Weak Magnetic Field  
$(\:\omega_c\lsim\: \omega_0\:)$}

Here, we focus on the properties of the spatial distribution of current
in the weak field region $(\:\omega_c\lsim\makebox[0.3em]{}\omega_0\:)$. 
At low temperature $(\:T \lsim\makebox[0.3em]{}\hbar\omega_0$, i.e. in ``MF''),
the magnetic moment shows a large fluctuation as a function of the field.
The spatial distribution of current in such a situation is
shown in Fig.~\ref{Current}(a), which indicates that
$J_{\theta}(r)$ can be either positive or negative, i.e. 
paramagnetic or diamagnetic, respectively.
In the bulk region ( mainly, $\:r\lsim\makebox[0.3em]{}R\:)$, 
large currents flow paramagnetically or diamagnetically depending on $r$.
This large bulk currents are sensitive to the strength of the field and
lead to the large fluctuation of magnetic moment.
This behavior of $J_{\theta}(r)$ is a characteristic feature of local currents
in the region of ``MF''.

As the temperature is raised $(\:T\gsim\makebox[0.3em]{}\hbar\omega_0\:)$
under such a weak field,
the fluctuation of magnetic moment disappears and the magnetic moment
shows the Landau diamagnetism as discussed in the previous section.
According with this change of magnetic moment, 
the spatial distribution of current $J_{\theta}(r)$ is changed as shown 
in Fig.~\ref{Current}(b);
the fluctuating large bulk currents are immediately reduced 
and finally the diamagnetic current flowing along the edge ($\:r\simeq R\:$) 
only survives.
It is seen that this property of the current distribution in ``LD'' is 
characteristic of two-dimensional confined system as deduced from the
Kubo's formula of the current distribution, which will be explained in the following.


Based on the Wigner representation, 
Kubo~\cite{Kubo} derived the analytic form of a current distribution 
proportional to a magnetic field and leading to 
the Landau diamagnetism in the system under a confining potential $V(r)$, 
which is assumed to be slowly varying in space compared to the electron wave length.
The expansion parameters in this theory are 
$\:\hbar^2 e \makebox[0.1em]{}H\makebox[0.1em]{} \frac{{\rm d}V(r)}{{\rm d}r}
/m^{3/2}c\makebox[0.2em]{} T^{5/2}\:$,
$\:\:(\hbar e H/m\makebox[0.1em]{} c\makebox[0.1em]{} T)^2\:$, 
$\:\:\hbar^{2}\left(\frac{{\rm d}V(r)}{{\rm d}r}\right)^{2}
/m\makebox[0.1em]{} T^{3}\:$ and
$\:\hbar^{2}\makebox[0.1em]{}\frac{{\rm d}^{2}V(r)}{{\rm d}r^{2}}
/m\makebox[0.1em]{} T^{2}\:$. 
In the present model of harmonic confining potential, the parameters 
are explicitly given as
$\:\:\omega_c/\omega_0\cdot(T/\hbar\omega_0)^{-5/2}
\cdot R/\xi\:$,
$\:\:(\omega_c/\omega_0)^2\cdot(T/\hbar\omega_0)^{-2}\:$,
$\:\:(T/\hbar\omega_0)^{-3}\cdot(R/\xi)^{2}\:$ and 
$\:(T/\hbar\omega_0)^{-2}\:$ respectively,
by replacing $r$ with the system radius $R$. After all,
$\:T/\hbar\omega_0\gsim\:\:(R/\xi)^{2/3}\:$ and 
$\:T/\hbar\omega_0\gsim\:\:\omega_c/\omega_0\:$
are required for the validity of the expansion in the Wigner representation,
which are actually a part of the region  ``LD'' in Fig.~\ref{Phase} 
( However, such an expansion turns out to be an asymptotic one 
since our analytic result shows that $\hbar=0$ is an essential singularity. 
See Appendix~\ref{Relation} for a detail. ).
The spatial distribution of current, which is valid under such a condition, 
is given as follows in general,
\begin{equation}  
\vecvar{J}(\vecvar{r})=
\frac{1}{3}\makebox[0.2em]{}c\makebox[0.1em]{}\mu_B^2 \makebox[0.2em]{}
\vecvar{H}\times\nabla n(\vecvar{r}),
\end{equation}  
where
\begin{equation}  
n(\vecvar{r})=\frac{1}{h^d}\int{\rm d}\vecvar{\pi}\int{\rm d}E
\makebox[0.3em]{}\delta \makebox[0.1em]{}' 
\left(E-\frac{\vecvar{\pi}^2}{2m}-V(\vecvar{r})\right)f(E),
\label{integral}
\end{equation} 
$d$ is the dimension of the system 
and $\vecvar{\pi}$ corresponds to a physical momentum.
The magnetic moment due to this current density becomes $\chi_L H$, 
the Landau diamagnetism.

In the present two-dimensional system, 
$\vecvar{\pi}$ and $\makebox[0.2em]{} E$-integration in eq.~(\ref{integral})
can be easily performed and $n(\vecvar{r})$ is given by
\begin{equation}
n(\vecvar{r})=\frac{2\pi m}{\hbar^2}f(V(\vecvar{r})),
\end{equation}  
and $\vecvar{J}(\vecvar{r})$ becomes
\begin{equation} 
\vecvar{J}(\vecvar{r})=
\frac{1}{3}\makebox[0.2em]{}c\makebox[0.1em]{}\mu_B^2 \cdot
\frac{2\pi m }{\hbar^2}\cdot
\vecvar{H}\times\nabla V(\vecvar{r})\cdot f\makebox[0.1em]{}'(V(\vecvar{r})).
\end{equation}  
Due to the factor $f\makebox[0.1em]{}'(V(\vecvar{r}))$
which reflects Fermi degeneracy,
this form of a current distribution 
means that the diamagnetic current flows
only in the region $V(\vecvar{r})\simeq \mu\makebox[0.2em]{}$, 
i.e. the edge of the system,
even if the potential is varying spatially in the bulk region.\\
Applying the above formula to the present model 
$V(r)=m\makebox[0.1em]{}\omega_0^2\makebox[0.1em]{} r^2/\makebox[0.1em]{}2$, 
we get $\vecvar{J}(\vecvar{r})$ as 
\begin{equation} 
\vecvar{J}(\vecvar{r})=-\makebox[0.2em]{}
\frac{j_0}{6}\cdot\frac{\omega_c}{\omega_0}\cdot
\left(\frac{T}{\hbar\omega_0}\right)^{-1}\cdot\frac{r}{\xi}\cdot
\frac{{\rm e}^{\beta(V(r)-\mu)}}
{(\makebox[0.2em]{}{\rm e}^{\beta(V(r)-\mu)}+1\makebox[0.2em]{})^2}\cdot
\vecvar{e}_{\theta},
\label{Wigner}
\end{equation}  
where 
\begin{equation} 
\beta(V(r)-\mu)=
\frac{1}{2}\cdot\left(\frac{T}{\hbar\omega_0}\right)^{-1}
\left[\left(\frac{r}{\xi}\right)^2-\left(\frac{R}{\xi}\right)^2\right],
\end{equation} 
and $j_0=e\makebox[0.1em]{}\omega_0/4\pi\makebox[0.1em]{}\xi$ 
the characteristic current density.
The current distribution calculated by eq.~(\ref{Wigner}) 
coincides with the one obtained 
from eq.~(\ref{Jr}). ( For example, at $\omega_c/\omega_0=0.1$ and 
$T/\hbar\omega_0=10$, the agreement is within $0.5\%$ numerically.)

In comparison, the case of $d=3$ which Kubo considered
appears somewhat different, 
since the current both at the surface region 
( the region as $\:V(\vecvar{r})\simeq\mu\:)$
and the bulk region can contribute to the magnetic moment.  
To see this, we approximate a Fermi distribution function in eq.~(\ref{integral})
by the linearized form around the Fermi energy as
\begin{eqnarray}
f(E)\simeq
\left\{ \begin{array}{ll}
\makebox[2em]{} 1 \makebox[3.8em]{}
(\makebox[0.5em]{}E<\mu-2\makebox[0.1em]{}T\makebox[0.5em]{})\\
\makebox[0.3em]{} \frac{1}{2}-\frac{E-\mu}{4\makebox[0.1em]{}T}\makebox[2.2em]{} 
(\makebox[0.5em]{}|\makebox[0.1em]{}E-\mu\makebox[0.1em]{}|\leq 2
\makebox[0.1em]{}T\makebox[0.5em]{})\makebox[1em]{}.\\
\makebox[1.8em]{}0\makebox[4em]{} 
(\makebox[0.5em]{}E>\mu+2\makebox[0.1em]{}T\makebox[0.5em]{})
\end{array}\right. 
\end{eqnarray}  
We assume the confining potential $V(\vecvar{r})$ is only dependent on $r$
in cylindrical coordinates $(r,\theta,z)$.
Then the current distribution is given by
\begin{equation}
\vecvar{J}(\vecvar{r})=-\frac{\gamma H}{\sqrt{T}}\cdot
V'(r)\cdot g(V(r))\makebox[0.3em]{}\vecvar{e}_{\theta},
\end{equation}
where $\gamma=\sqrt{2}\pi m^{3/2}c\mu_B^{2}/3h^{3}$ and
$g(V(r))$, the function corresponding to $f'(V(r))$ in the case of $d=2$ is given as
\begin{eqnarray}
g(V(r))=
\left\{ \begin{array}{ll}
\makebox[0em]{} 
\sqrt{\frac{\mu-V(r)}{T}+2}-\sqrt{\frac{\mu-V(r)}{T}-2}
\makebox[2.8em]{}
(\makebox[0.5em]{}V(r)<\mu-2\makebox[0.1em]{}T\makebox[0.5em]{})\\[0.5em]
\makebox[3.5em]{} 
\sqrt{\frac{\mu-V(r)}{T}+2}
\makebox[6em]{} 
(\makebox[0.5em]{}|\makebox[0.2em]{}V(r)-\mu\makebox[0.2em]{}|\leq 2
\makebox[0.1em]{}T\makebox[0.5em]{})
\makebox[1em]{}.\\[0.5em]
\makebox[6.5em]{}0\makebox[8.1em]{} 
(\makebox[0.5em]{}V(r)>\mu+2\makebox[0.1em]{}T\makebox[0.5em]{})
\end{array}\right. 
\end{eqnarray} 
In the bulk region where $\:V(r)\ll \mu-2\makebox[0.1em]{}T\:$,
$\:g(V(r))$ becomes
\begin{equation}
g(V(r))\simeq
\frac{2}{\sqrt{\frac{\mu-V(r)}{T}}}\ll 1\makebox[1em]{},
\end{equation} 
while  at the surface region,
$|\makebox[0.1em]{}V(r)-\mu\makebox[0.1em]{}|<2\makebox[0.1em]{}T\:$,
$\:g(V(r))\sim 1\:$.
Therefore, it can be said that
the current causing the Landau diamagnetism
is mainly induced at the surface also in a three-dimensional system,
which is not so clear as the case of $\:d=2$.   

At low temperature as $\:T\lsim\makebox[0.3em]{}\hbar\omega_0\:$, 
on the other hand, the contributions of higher order terms in magnetic field 
become larger
and the above formula of current density is not valid.
Then, the current distribution
changes dramatically at around the temperature $T=\hbar\omega_0$
under a weak field $(\:\omega_c\lsim\makebox[0.3em]{}\omega_0\:)$
as is shown in Fig.~\ref{Current}(a) and (b) and
this change is clearly reflected in a magnetic moment of the system. 
Hajdu and Shapiro~\cite{Shapiro} 
studied a two-dimensional system under a confining potential 
$V(\vecvar{r})=m\omega_0^2 x^2/2$ ( i.e. harmonic groove ) with a width $L_x$ 
and an arbitrary long length $L_y$ by imposing a periodic boundary condition in the 
$y$-direction, and pointed out that
the temperature as $T=\hbar\omega_0$ below which magnetic moment shows a large 
fluctuation under a weak field $(\:\omega_c\lsim\:\:\omega_0\:)$
corresponds to $\hbar/\tau_{tr}$ where $\tau_{tr}=L_x/v_F$,
time of propagation for electrons at the Fermi energy across the groove.    
The physical reason why the field dependence of 
magnetic moment shows such a dramatic difference depending on the temperature is
understood as follow. 
When the effect of a confining potential $V(\vecvar{r})$ is considered perturbative,
the thermal Green's function of a degenerate electrons has the 
following damping factor,
\begin{equation}
G(x,\tau=0)\makebox[0.3em]{} \propto \makebox[0.5em]{}
{\rm e}^{-\frac{\pi T}{\hbar v_F}x}.
\end{equation}
Therefore, the length $\hbar v_F/\pi T\equiv l_c$ is regarded as the coherence length
of degenerate electrons.
Under the condition $l_c\gsim\makebox[0.3em]{} L$, where $L$ is a system length,
electrons near the Fermi surface can propagate 
from one side of the system to the 
other side with a small damping and experience the multiple reflection
by the boundary potential as shown in Fig.~{\ref{reflection}}(a).
Therefore electrons near the Fermi surface, 
which play an essential role in the orbital magnetism of degenerate electron
system, are strongly affected by the boundary potential. 
In this model, 
the condition $l_c\gsim\makebox[0.3em]{} L\makebox[0.3em]{}$ corresponds to 
$\:T\lsim\makebox[0.3em]{}\hbar\omega_0\makebox[0.3em]{}$ i.e. ``MF''.
As shown in Fig.~\ref{Current} (a), it is seen that
this multiple reflection induces large currents irregularly distributed
paramagnetically or diamagnetically in the bulk region, 
which are sensitive to the strength of a magnetic field and causes
the large fluctuation of a magnetic moment as a function of the field.
On the other hand, under the condition 
$l_c\lsim\makebox[0.3em]{} L\makebox[0.3em]{}$
$(\:T\gsim\makebox[0.3em]{}\hbar\omega_0\:)$,
the effect of the multiple reflection by the boundary potential wall is reduced
as shown in Fig.~\ref{reflection}(b),
and this is considered to lead to
the suppression of the bulk currents and the recovery
of the Landau diamagnetism.
Considering the case of a harmonic groove system  based on this idea,
which Hajdu and Shapiro~\cite{Shapiro} studied,
it is natural that the relative magnitude of $l_c$ and the width $L_x$ 
affects the field dependence of magnetic moment while $L_y$ does not affect it 
qualitatively, since the $x$-direction is actually confined by a harmonic potential
and a multiple reflection can happen only in the $x$-direction. 
 
Robnik~\cite{Robnik} discussed the size effect on the zero-field susceptibility
in a system confined by a hard wall,
based on the Green's function method, and
concluded that at $T=0$
the contribution of the boundary wall to the susceptibility is always paramagnetic
but with the order of magnitude $(k_F L)^{-1}$ 
compared to the Landau susceptibility, where $k_F$ is a Fermi wave number
and $L$ is a system length.
As is clear from Fig.~\ref{Moment}(a),
this evaluation of a size effect due to the confining potential
is too small.
This disagreement can be attributed to
the fact that the multiple reflection by the boundary potential wall is disregarded
in ref. 9.


\subsection{Region of Strong Magnetic Field  
$(\:\omega_c\gsim\:\omega_0\:)$}

Under a strong magnetic field
$(\makebox[0.3em]{}\omega_c\gsim\makebox[0.3em]{} \omega_0\makebox[0.3em]{})$,
the energy spectrum in eq.~(\ref{energy}) becomes close to the one 
without the confining potential, i.e. the Landau level 
( the states characterized by $n$ in eq.~(\ref{energy}) ).
At low temperature as
$T\lsim\makebox[0.3em]{}\hbar\omega_c$,
magnetic moment is affected by this Landau quantization and 
oscillates as a function of the field as is shown in Fig.~\ref{Moment}(c) and (d)
(This is the familiar de Haas-van Alphen effect).
In this ``dHvA'', the spatial distribution of current is as is shown in 
Fig.~\ref{Current}(c).
Several diamagnetic peaks can be seen and paramagnetic currents flow between them.
These bulk currents are much larger than the edge current in ``LD'', although
the diamagnetic currents are almost cancelled by the paramagnetic ones as an average. 
Each diamagnetic peak comes from the state with the same $n$ in eq.~(\ref{energy})
which corresponds to Landau level.
In this model, each Landau level has
a different degeneracy with respect to the angular momentum, $\alpha$, 
due to the confining potential and the lower Landau level has the higher degeneracy.
Therefore, the spatial extent of the lower Landau level with such a  
degeneracy becomes larger and is reflected in the current distribution 
as the spatially separated diamagnetic peaks.
As the field is raised, the number of Landau level below the Fermi energy
decreases and the diamagnetic peaks 
closest to the center of the system disappear one by one. This change of 
the current distribution causes the large oscillation of the magnetic moment as a 
function of the field.
At much lower temperature in ``MF'' 
$(\:T\lsim\makebox[0.3em]{}\hbar\omega_{-}\:)$
where the AB effect appears, the current distribution
is almost same as in ``dHvA'' region.

At higher temperature as $T\gsim\makebox[0.3em]{}\hbar\omega_c$,
the diamagnetic peaks in the bulk region
are smeared out and there remain only edge current as is shown
in Fig.~\ref{Current}(d). 
This edge current is spatially broadened compared to 
the one at lower temperature in ``LD'' 
shown in Fig.~\ref{Current}(b).
This temperature dependence of edge current is understood from eq.~(\ref{Wigner}) 
where the spatial extent of edge current corresponds to  
$|\makebox[0.1em]{}V(r)-\mu\makebox[0.1em]{}|\lsim\makebox[0.3em]{}T$.

\section{Summary}
\label{L4}
We have studied the relation between the spatial distribution of current
and the magnetic moment of the whole system
in a two-dimensional electron system under an isotropic harmonic potential as
$\:V(\vecvar{r})=m\makebox[0.1em]{}\omega_0^2\makebox[0.1em]{}r^2/\makebox[0.1em]{}2\:$.
It is found that 
characteristic dependences of magnetic moment on temperature and magnetic field
are clearly reflected in the spatial distribution of current.

Under a weak field $(\:\omega_c\lsim\makebox[0.3em]{}\omega_0\:)$,
the field dependence of magnetic moment dramatically changes at around the temperature 
$\:T=\hbar\omega_0\:$.
In the low temperature region $(\:T\lsim\makebox[0.3em]{}\hbar\omega_0\:)$,
magnetic moment shows a large fluctuation as a function of the field,
as was indicated by Yoshioka and Fukuyama~\cite{DY}.
We attribute this large fluctuation of magnetic moment 
to a multiple reflection by the boundary confining potential,
which becomes important once the coherence length of degenerate electrons
$\:l_c=\hbar v_F/\pi T\:$ gets longer than a system length.
In the present model, the system length is characterized by the radius $R$ and
the above condition leads to $l_c\gsim\makebox[0.3em]{}R$ implying 
$\:T\lsim\makebox[0.3em]{}\hbar\omega_0\:$.  
At such low temperature, it is seen that 
the multiple reflection induces 
large currents irregularly distributed paramagnetically or diamagnetically 
in the bulk region, which are sensitive to the strength of
the field and cause a large fluctuation of the magnetic moment.
As the temperature is raised $(\:T\gsim\makebox[0.3em]{}\hbar\omega_0\:)$,
the fluctuations of the magnetic moment
are reduced and the usual Landau diamagnetism is recovered.
Corresponding to this change, it is found that
the large currents in the bulk region are immediately reduced and 
finally the diamagnetic current flowing along the edge 
( in the region satisfying  $V(\vecvar{r})\simeq\mu$ ) only survives,
which leads to the Landau diamagnetism. 
This edge current is regarded characteristic of a two-dimensional confined system,
as inferred from the Kubo~\cite{Kubo}'s formula 
of the diamagnetic current distribution 
which is derived for high temperature as 
$\:T\gsim\makebox[0.3em]{}\hbar\omega_0\cdot (R/\xi)^{2/3}\:$
and $\:\hbar\omega_c\:$, 
where $\:\xi=\sqrt{\hbar/m\omega_0}\:$, the characteristic length.
From this formula, the cancellation of the bulk currents is seen to be due to 
the Fermi degeneracy.
It is also noted that the persistence of the diamagnetic current at the surface 
is also seen in a three-dimensional system.  

Under a strong field $(\:\omega_c\gsim\makebox[0.3em]{}\omega_0\:)$,
dHvA effect appears at low temperature as 
$\:T\lsim\makebox[0.3em]{}\hbar\omega_c\:$.
In this situation, the current is distributed in the bulk region and 
diamagnetic peaks appear, although 
paramagnetic currents flow between the peaks,
which almost cancel the contribution of the diamagnetic peaks as an average.
Here, each peak is due to 
the Landau level with different degeneracy because of the confining potential.
As the field is increased,  
this diamagnetic peak disappears one by one
corresponding to the decrease of the number of Landau levels
under the Fermi energy, and this causes
the large oscillation of magnetic moment (dHvA effect).
At much lower temperature 
$(\:T\lsim\makebox[0.3em]{}\hbar\omega_{-}
\simeq\makebox[0.3em]{}\hbar\omega_0^2/\omega_c\:$, the energy spacing between 
different angular momentum states at each Landau level ), 
the small but rapid oscillation 
with a period $\phi_0=hc/e$ appears as a function of the total flux $\phi$ 
in addition to the dHvA oscillation, , 
which was originally found by Meir et al~\cite{NA} at $T=0$.
This oscillation is caused by a coherent motion of electrons along the edge
and can be called AB oscillation.
On the other hand, at high temperature as $T\gsim\:\:\hbar\omega_c$ 
under such a strong field,
the bulk currents are quickly reduced and the Landau diamagnetism by the edge current
is recovered as in the case under a weak field.
It should be noticed that this edge current 
does not cause AB oscillations, because the current is due to incoherent motions
of electrons.

\section*{Acknowledgment}
One of the authors (Y. I) would like to express 
his sincerest gratitude to Hiroshi Kohno 
for stimulating and instructive discussions
and thank 
Masakazu Murakami for instructive and useful discussions.

\appendix
\section{Derivation of Magnetic Moment from Thermodynamical Potential $\Omega$}
\label{LA}
The thermodynamic potential $\Omega$ is given by
\begin{equation}
\Omega=-\frac{1}{\beta}
\sum_{n=0}^{\infty}\sum_{\alpha=-\infty}^{\infty}
\log\left[\makebox[0.2em]{}1+{\rm e}^{-\beta(E_{n \alpha}-\mu)}\makebox[0.2em]{}\right],
\end{equation}
where $E_{n \alpha}=\hbar\omega(n+1/2)+\hbar\omega|\alpha|/2+\hbar\omega_c
\makebox[0.2em]{}\alpha/2$.\\
By applying the Poisson summation formula 
\begin{equation}
\sum_{n=0}^{\infty}F\left(n+\frac{1}{2}\right)=
\int_{0}^{\infty}{\rm d}x \makebox[0.2em]{} F(x)
+2\makebox[0.2em]{}{\rm Re}\sum_{k=1}^{\infty}(-1)^k
\int_{0}^{\infty}{\rm d}x \makebox[0.2em]{}F(x) 
\makebox[0.2em]{}{\rm e}^{2\pi {\rm i} kx},
\end{equation}
to the sum over $n$, $\Omega$ is transformed as
\begin{eqnarray}
\Omega =&-&\frac{1}{\beta\hbar\omega}\int_{0}^{\infty}{\rm d}\varepsilon\makebox[0.2em]{}
\left[\makebox[0.2em]{}
1+2\makebox[0.2em]{}{\rm Re}\sum_{k=1}^{\infty}(-1)^k 
{\rm e}^{2\pi{\rm i}k\frac{\varepsilon}{\hbar\omega}}
\right]
\log\left[\makebox[0.2em]{}1+{\rm e}^{-\beta(\varepsilon-\mu)}\right]\nonumber\\
&-&\frac{1}{\beta\hbar\omega}\int_{0}^{\infty}{\rm d}\varepsilon
\left[\makebox[0.2em]{}
1+2\makebox[0.2em]{}{\rm Re}\sum_{k=1}^{\infty}(-1)^k 
{\rm e}^{2\pi{\rm i}k\frac{\varepsilon}{\hbar\omega}}
\right]
\sum_{\sigma=\pm}\sum_{\alpha=1}^{\infty}
\log\left[\makebox[0.2em]{}1
+{\rm e}^{-\beta(\hbar\omega_{\sigma}\alpha+\varepsilon-\mu)}\right],
\end{eqnarray}
where $\omega_{\pm}=(\omega\pm\omega_c)/2$.\\
To the sum over $\alpha$, we again use the Poisson summation formula as
\begin{equation}
\frac{1}{2}\makebox[0.2em]{}F(0)+\sum_{\alpha=1}^{\infty}F\left(\alpha\right)=
\int_{0}^{\infty}{\rm d}x \makebox[0.2em]{}F(x)
+2\makebox[0.2em]{}{\rm Re}\sum_{l=1}^{\infty}
\int_{0}^{\infty}{\rm d}x \makebox[0.2em]{}F(x)
\makebox[0.2em]{} {\rm e}^{2\pi {\rm i} lx}.
\end{equation}
Then $\Omega$ is given as
\begin{eqnarray}
\Omega &=&-\frac{1}{\beta(\hbar\omega_0)^2}
\int_{0}^{\infty}{\rm d}\varepsilon
\int_{0}^{\infty}{\rm d}\eta
\left[\makebox[0.2em]{}
1+2\makebox[0.2em]{}{\rm Re}\sum_{k=1}^{\infty}(-1)^k 
{\rm e}^{2\pi{\rm i}k\frac{\varepsilon}{\hbar\omega}}
\right]
\cdot\log\left[\makebox[0.2em]{}1+{\rm e}^{-\beta(\varepsilon+\eta-\mu)}\right]\nonumber\\
&&-\sum_{\sigma=\pm}
\frac{2}{\beta\hbar\omega\cdot\hbar\omega_{\sigma}}
\makebox[0.2em]{}{\rm Re}\sum_{l=1}^{\infty}
\int_{0}^{\infty}{\rm d}\varepsilon
\int_{0}^{\infty}{\rm d}\eta
\makebox[0.2em]{}
\log\left[\makebox[0.2em]{}1+{\rm e}^{-\beta(\varepsilon+\eta-\mu)}\right]\cdot
{\rm e}^{2\pi{\rm i}l\frac{\eta}{\hbar\omega_{\sigma}}}\\
&&-\sum_{\sigma=\pm}
\frac{4}{\beta\hbar\omega\cdot\hbar\omega_{\sigma}}
\makebox[0.2em]{}{\rm Re}\makebox[0.2em]{}\sum_{k=1}^{\infty}(-1)^k
\int_{0}^{\infty}{\rm d}\varepsilon\makebox[0.2em]{}
{\rm e}^{2\pi{\rm i}k\frac{\varepsilon}{\hbar\omega}}\cdot
{\rm Re}\sum_{l=1}^{\infty}
\int_{0}^{\infty}{\rm d}\eta\makebox[0.2em]{}
\makebox[0.2em]{}\log\left[\makebox[0.2em]{}1
+{\rm e}^{-\beta(\varepsilon+\eta-\mu)}\right]\cdot
{\rm e}^{2\pi{\rm i}l\frac{\eta}{\hbar\omega_{\sigma}}}.\nonumber
\end{eqnarray}
We perform integrations over $\varepsilon$ and $\eta$ twice by parts,
\begin{eqnarray} 
&&\int_{0}^{\infty}{\rm d}\varepsilon\makebox[0.2em]{}
\log\left[\makebox[0.2em]{}1+{\rm e}^{-\beta(\varepsilon+\eta-\mu)}\right]\cdot
{\rm e}^{2\pi{\rm i}k\frac{\varepsilon}{\hbar\omega}}\nonumber\\
&=&
-\frac{\hbar\omega}{2\pi{\rm i}k}\log\left[\makebox[0.2em]{}1
+{\rm e}^{-\beta(\eta-\mu)}\right]
+\beta\left(\frac{\hbar\omega}{2\pi k}\right)^2 f(\eta)\nonumber\\
[1.5mm]
&&-\beta\left(\frac{\hbar\omega}{2\pi k}\right)^2
{\rm e}^{2\pi{\rm i}k\frac{\mu-\eta}{\hbar\omega}}
\int_{-\beta(\mu-\eta)}^{\infty}{\rm d}\xi\makebox[0.2em]{}
\frac{{\rm e}^{\xi}}{({\rm e}^{\xi}+1)^2}\cdot
{\rm e}^{{\rm i}\frac{2\pi k}{\beta \hbar\omega}\xi}.
\end{eqnarray}
In the last integration term, we assume $\mu\gg T$ and approximate as
\begin{eqnarray} 
&&\int_{-\beta(\mu-\eta)}^{\infty}{\rm d}\xi\makebox[0.2em]{}
\frac{{\rm e}^{\xi}}{({\rm e}^{\xi}+1)^2}\cdot
{\rm e}^{{\rm i}\frac{2\pi k}{\beta \hbar\omega}\xi}\nonumber\\
&\simeq&
\int_{-\infty}^{\infty}{\rm d}\xi\makebox[0.2em]{}
\frac{{\rm e}^{\xi}}{({\rm e}^{\xi}+1)^2}\cdot
{\rm e}^{{\rm i}\frac{2\pi k}{\beta \hbar\omega}\xi}
\cdot \theta\makebox[0.1em]{}[\makebox[0.1em]{}\mu-\eta\makebox[0.1em]{}]\nonumber\\
&=&\frac{\frac{2\pi^2 k}{\beta\hbar\omega}}
{\sinh\left(\frac{2\pi^2 k}{\beta\hbar\omega}\right)}
\cdot \theta\makebox[0.1em]{}[\makebox[0.1em]{}\mu-\eta\makebox[0.1em]{}],
\end{eqnarray} 
where $\theta\makebox[0.1em]{}[\makebox[0.1em]{}x\makebox[0.1em]{}]$ 
is the Heaviside function.\\
By using this approximation and the relations
\begin{eqnarray}
\sum_{k=1}^{\infty}\frac{(-1)^{k+1}}{k^2}=\frac{\pi^2}{12}\makebox[0.2em]{},
\makebox[1em]{}\sum_{k=1}^{\infty}\frac{1}{k^2}=\frac{\pi^2}{6}\makebox[0.2em]{},
\end{eqnarray}
we get the form represented in eq.~(\ref{EachOmega}).

\section{Derivation of Magnetic Moment from Current Distribution}
\label{LB}
By replacing $r^{2}/2l^{2}$ in eq. (\ref{MJ}) with $\rho$, 
magnetic moment is given by
\begin{eqnarray}
M=-\mu_B\sum_{n=0}^{\infty}\sum_{\alpha=-\infty}^{\infty}
\frac{n\makebox[0.2em]{}!}
{(\makebox[0.2em]{}n+|\alpha|\makebox[0.2em]{})\makebox[0.2em]{}!}
\makebox[0.2em]{}f(E_{n\alpha})\cdot
\int_{0}^{\infty}{\rm d}\rho\makebox[0.2em]{} 
\left[\makebox[0.2em]{}\alpha+\frac{\omega_c}{\omega}\rho\makebox[0.2em]{}\right]
{\rm e}^{-\rho}\rho^{|\alpha|}{L_{n}^{(|\alpha|)}[\rho]}^2.
\end{eqnarray}
$\rho$-integrations can be performed as
\begin{eqnarray}
\int_{0}^{\infty}{\rm d}\rho \makebox[0.4em]{}
{\rm e}^{-\rho}\rho^{|\alpha|}\makebox[0.2em]{}L_{n}^{(|\alpha|)}[\rho]
\makebox[0.2em]{}L_{m}^{(|\alpha|)}[\rho]=
\frac{(\makebox[0.2em]{}n+|\alpha|\makebox[0.2em]{})\makebox[0.2em]{}!}
{n\makebox[0.2em]{}!}\makebox[0.2em]{}\delta_{n m}
\makebox[0.2em]{},
\end{eqnarray}
and
\begin{eqnarray}
&&\int_{0}^{\infty}{\rm d}\rho\makebox[0.4em]{} 
{\rm e}^{-\rho}\rho^{|\alpha|+1}{L_{n}^{(|\alpha|)}[\rho]}^2 \nonumber\\
&=&
\int_{0}^{\infty}{\rm d}\rho\makebox[0.4em]{} 
{\rm e}^{-\rho}\rho^{|\alpha|+1}
\left[\makebox[0.2em]{}L_{n}^{(|\alpha|+1)}[\rho]
-L_{n-1}^{(|\alpha|+1)}[\rho]\makebox[0.2em]{}\right]^2\nonumber\\
&=&\frac{(\makebox[0.2em]{}n+|\alpha|+1\makebox[0.2em]{})\makebox[0.2em]{}!}
{n\makebox[0.2em]{}!}
+\frac{(\makebox[0.2em]{}n+|\alpha|\makebox[0.2em]{})\makebox[0.2em]{}!}
{(\makebox[0.1em]{}n-1\makebox[0.1em]{})\makebox[0.2em]{}!}\nonumber\\
&=&\frac{(\makebox[0.2em]{}n+|\alpha|\makebox[0.2em]{})\makebox[0.2em]{}!}
{n\makebox[0.2em]{}!}
\cdot(\makebox[0.2em]{}2n+|\alpha|+1\makebox[0.2em]{})\makebox[0.2em]{}.
\end{eqnarray}
Therefore, the magnetic moment becomes
\begin{eqnarray}
M&=&-\mu_B\sum_{n=0}^{\infty}\sum_{\alpha=-\infty}^{\infty}
\left[\makebox[0.2em]{}\alpha+\frac{\omega_c}{\omega}\cdot
\left(\makebox[0.2em]{}2n+|\alpha|+1\makebox[0.2em]{}\right)\makebox[0.1em]{}\right]
f(E_{n \alpha})\nonumber\\
&=&\sum_{n \alpha}
\left(-\frac{\partial E_{n \alpha}}{\partial H}\right)f(E_{n \alpha})\makebox[0.2em]{}.
\end{eqnarray}
This coincides with the one derived from the thermodynamical potential 
in eq.~(\ref{Statistic}).

\section{ Wigner Representation for Landau Diamagnetism }
\label{Relation}

In the present model of the harmonic confining potential,
the parameter region, where the Kubo's formula of current distribution leading to the 
Landau diamagnetism\cite{Kubo} is valid, is  
$\:T/\hbar\omega_0\gsim\:\:(R/\xi)^{2/3}=(2\cdot\mu/\hbar\omega_0)^{1/3}\:$ and 
$\:T/\hbar\omega_0\gsim\:\:\omega_c/\omega_0\:$,
which is dependent on the chemical potential, $\mu$, i.e. the number of electrons.
This region is only a small part of  ``LD'' region 
$(\:T/\hbar\omega_0\gsim\:\:1\:$ and $\:T/\hbar\omega_0\gsim\:\:\omega_c/\omega_0\:)$ of Fig.~\ref{Phase}.
The reason of this apparent discrepancy is as follows.

By differentiating the trigonometric functions in eq.~(\ref{osc})
with respect to the field, it is found that
$\Omega_{osc}$ contributes to the magnetic moment
in the order of magnitude 
$(2\pi^2 T/\hbar\omega)/\sinh (2\pi^2 T/\hbar\omega)$ or 
$(2\pi^2 T/\hbar\omega_{\pm})/\sinh (2\pi^2 T/\hbar\omega_{\pm})$
relative to $\Omega_{L}$, and
based on this fact the phase diagram in Fig.~\ref{Phase} is decided.
These factors $(2\pi^2 T/\hbar\omega)/\sinh (2\pi^2 T/\hbar\omega)$ and
$(2\pi^2 T/\hbar\omega_{\pm})/\sinh (2\pi^2 T/\hbar\omega_{\pm})$
can not be expanded in terms of $\hbar\omega/T$ and $\hbar\omega_{\pm}/T$,
which are seen to correspond to
the expansion parameters in the treatment of Wigner representation. 
This implies that
one cannot conclude whether the bulk magnetic moment shows the Landau diamagnetism or not,
unless the contributions of higher order terms in $\hbar$
in Wigner representation are summed up 
to the infinite order.
Actually, in the parameter region, 
$\:1\lsim\:\:T/\hbar\omega_0\lsim\:\:(R/\xi)^{2/3}$ and
$\:T/\hbar\omega_0\gsim\:\:\omega_c/\omega_0\:$, 
some expansion parameters in ref.~4 get larger than $1$, but
even in such a case, the Landau diamagnetism is realized
and diamagnetic current is mainly induced at the edge
as seen in Fig.~\ref{Current}(b).

\newpage
\begin{figure}
\caption{
Phase diagram showing characteristic regions of the field dependence 
of magnetic moment at various temperatures.
}
\label{Phase}
\end{figure}

\begin{figure}
\caption
{
Magnetic field dependences of magnetic moment at various temperatures.
The dependences under a weak field $(\:\omega_c\lsim\:\:\omega_0\:)$
are shown in (a) and (b), where
(a) corresponds to ``MF'' and (b) ranges from ``MF'' to ``LD''.
On the other hand, the dependences under a strong field
$(\:\omega_c\gsim\:\:\omega_0\:)$ are shown in (c) and (d), where
(c) and (d) range from ``MF'' to ``dHvA'' and from ``dHvA'' to ``LD'', respectively.
The number of electrons is fixed to $5000$.
}
\label{Moment}
\end{figure}

\begin{figure}
\caption{
A schematic illustration of the current flowing in the system.
$J_{\theta}(r)$ is defined positive in the direction as here.
}
\label{System}
\end{figure}

\begin{figure}
\caption{
Spatial distribution of current at various temperatures.
Here, (a) and (b) are the current distribution
under a weak field $(\:\omega_c/\omega_0=0.1\:)$, where
(a) corresponds to ``MF'' and (b) ranges from ``MF'' to ``LD''. 
(c) and (d) are the current distribution under  
a strong field $(\:\omega_c/\omega_0=20\:)$, where
(c) and (d) range from ``MF'' to ``dHvA''  and
from ``dHvA'' to ``LD'', respectively. 
The number of electrons is fixed to $5000$.
$J_{\theta}(r)$ is normalized by $j_0=e\omega_0
/4\pi\xi$.
}
\label{Current}
\end{figure}

\begin{figure}
\caption{
A schematic illustration of the reflection at the boundary confining potential;
(a) $\:l_c(=\hbar v_F /\pi T)\gsim\:\:L$ where
the multiple reflection becomes dominant, and
(b) $\:l_c\lsim\:\:L$ where
the boundary affects the behaviors of electrons only around the boundary.
}
\label{reflection}
\end{figure}


\begin{thebibliography}{99}
\bibitem{Bohr} N. Bohr: Ph.D. thesis, Univ. of Copenhagen (1911)
\bibitem{vanVleck}J. H. van Vleck: {\it The Theory of Electric and
Magnetic Susceptibilities}
(Clardendon Press, Oxford, 1932) \S 26, 81.
\bibitem{Landau1}L. D. Landau: Z. Phys. {\bf 64} (1930) 629.
\bibitem{Kubo}R. Kubo: J. Phys. Soc. Jpn. {\bf 19} (1964) 2127.
\bibitem{DY5}L. Friedman: Phys. Rev. {\bf 134} (1964) A336.
\bibitem{DY6}D. Childers and P. Pincus: Phys. Rev. {\bf 177} (1969) 1036.
\bibitem{DY7}R. V. Denton: Z. Phys. {\bf 265} (1973) 119.
\bibitem{DY8}F. Meir and P. Wyder: Phys. Rev. Lett. {\bf 30} (1973) 181.
\bibitem{DY9}A. I. Buzdin, O. V. Dolgov and Yu. E. Lozovik: 
Phys. Lett. {\bf 100A} (1984) 261.
\bibitem{Robnik}M. Robnik: J. Phys. A: Gen. Phys. {\bf 19} (1986) 3619.
\bibitem{DY12}R. Nemeth: Z. Phys. B {\bf 81} (1990) 89.
\bibitem{DY13}J. M. van Ruitenbeek and D. A. van Leeuwen: 
Phys. Rev. Lett. {\bf 67} (1991) 640.
\bibitem{NA}Y. Meir, O. Entin-Wohlmann and Y. Gefen: 
Phys. Rev. B {\bf 42} (1990) 8351.
\bibitem{DY}D. Yoshioka and H. Fukuyama: J. Phys. Soc. Jpn. 
{\bf 61} (1992) 2368.
\bibitem{Shapiro}J. Hajdu and B. Shapiro: Europhys. Lett {\bf 28} (1994) 61.
\bibitem{DY17}O. D. Cheishvili: Pis'ma Zh. Eksp. Theor. Fiz. {\bf 48} (1988)
206. translation: JETP Lett. {\bf 48} (1988) 225.
\bibitem{HF}H. Fukuyama: J. Phys. Soc. Jpn. {\bf 58} (1989) 47.
\bibitem{DY18}R. A. Serota and S. Oh: Phys. Rev. B {\bf 41} (1990) 10523.
\bibitem{DY19}S. Oh, A. Yu. Zyuzin and R. A. Serota: 
Phys. Rev. B {\bf 44} (1991) 8858.
\bibitem{HY}H. Yoshioka: Ph.D. thesis, Univ. of Tokyo (1992). 
H. Yoshioka and H. Fukuyama: 
{\it Transport Phenomena in Mesoscopic Systems, 
Springer Series of Solid State Sciences}, ed. H. Fukuyama and T. Ando 
(Springer, Berlin, 1992) 263.
\bibitem{Landau2}L. D. Landau and E. M. Lifshitz: 
{\it Statistical Physics, Part 1} (Pergamon, Oxford, 1980) \S 60. 
\end{thebibliography}
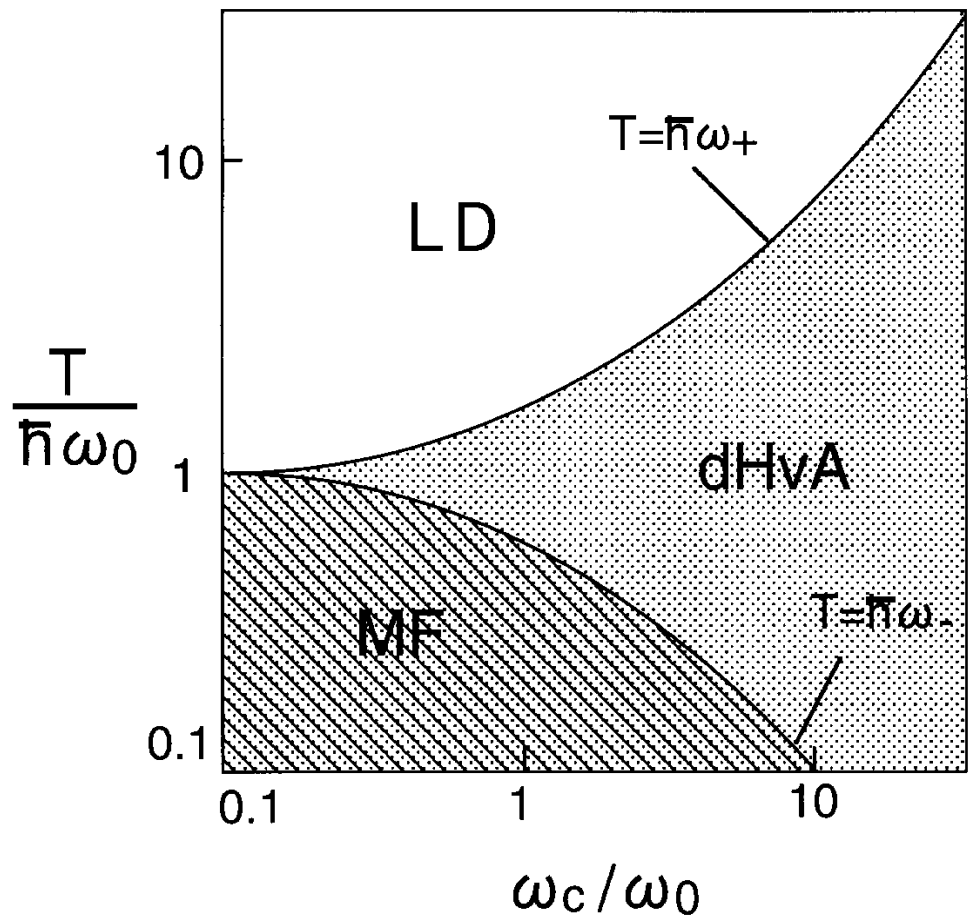
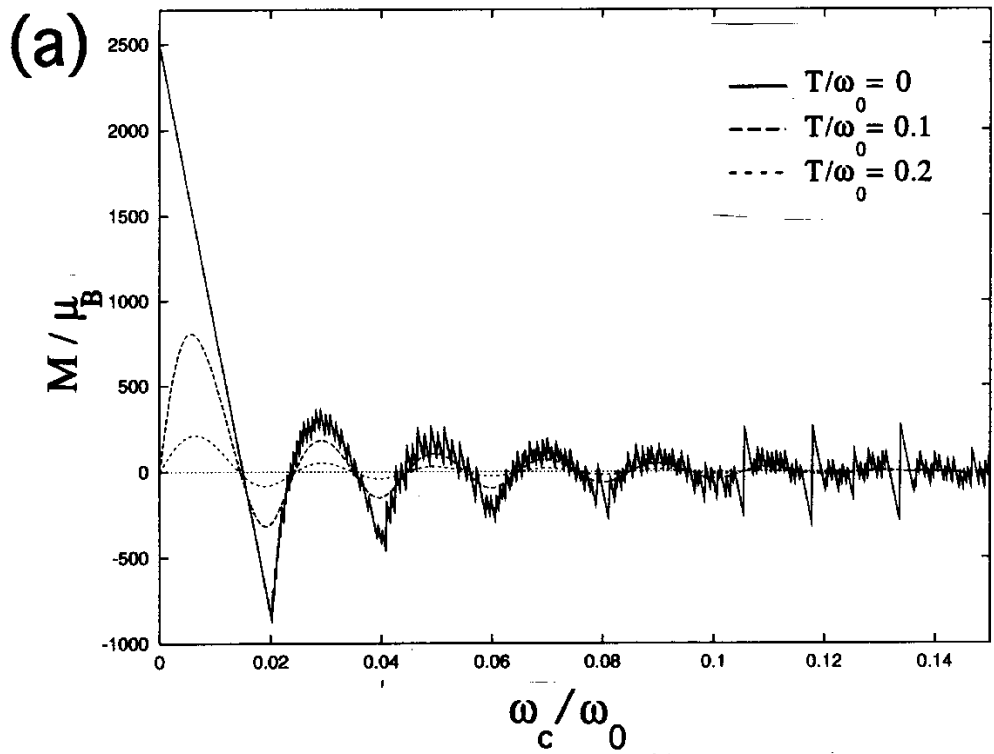
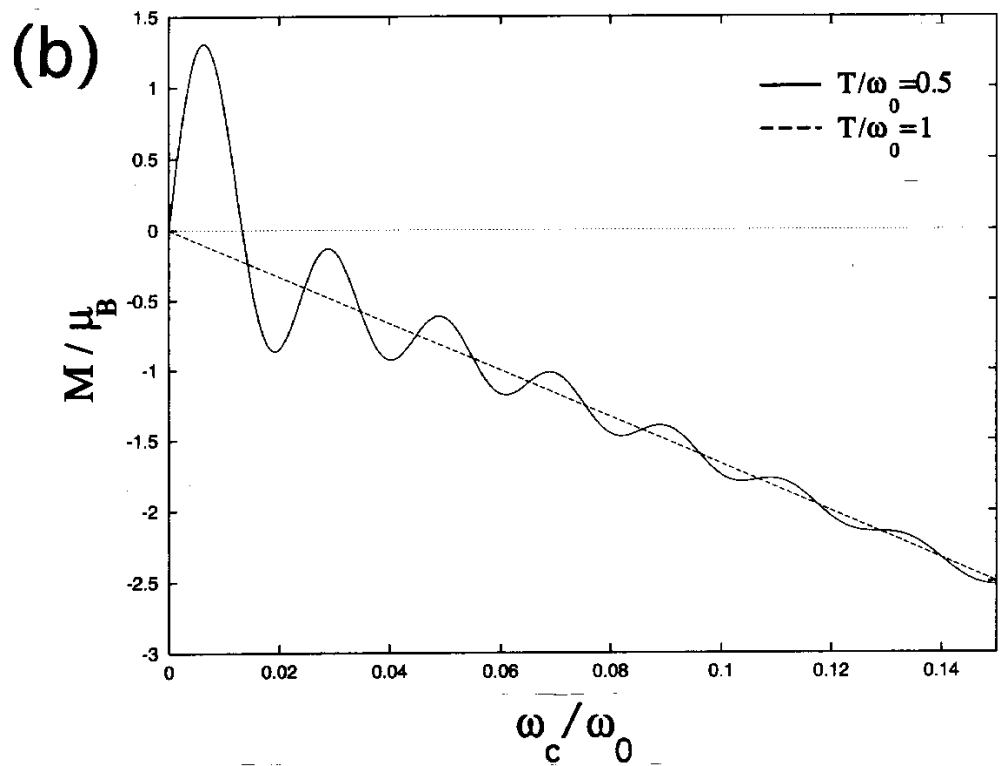
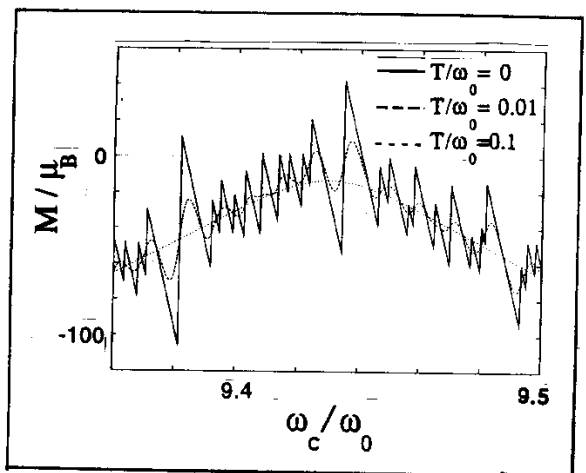
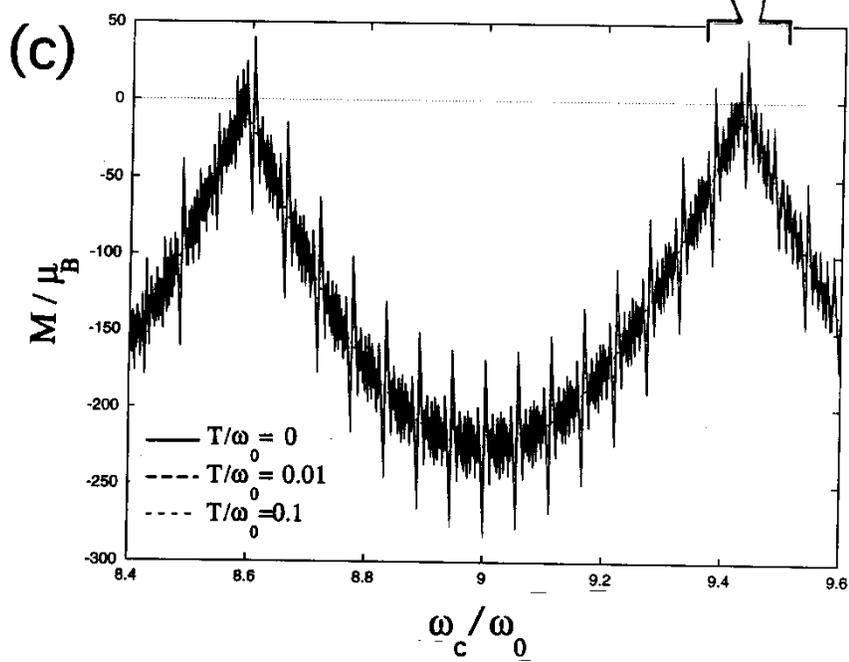
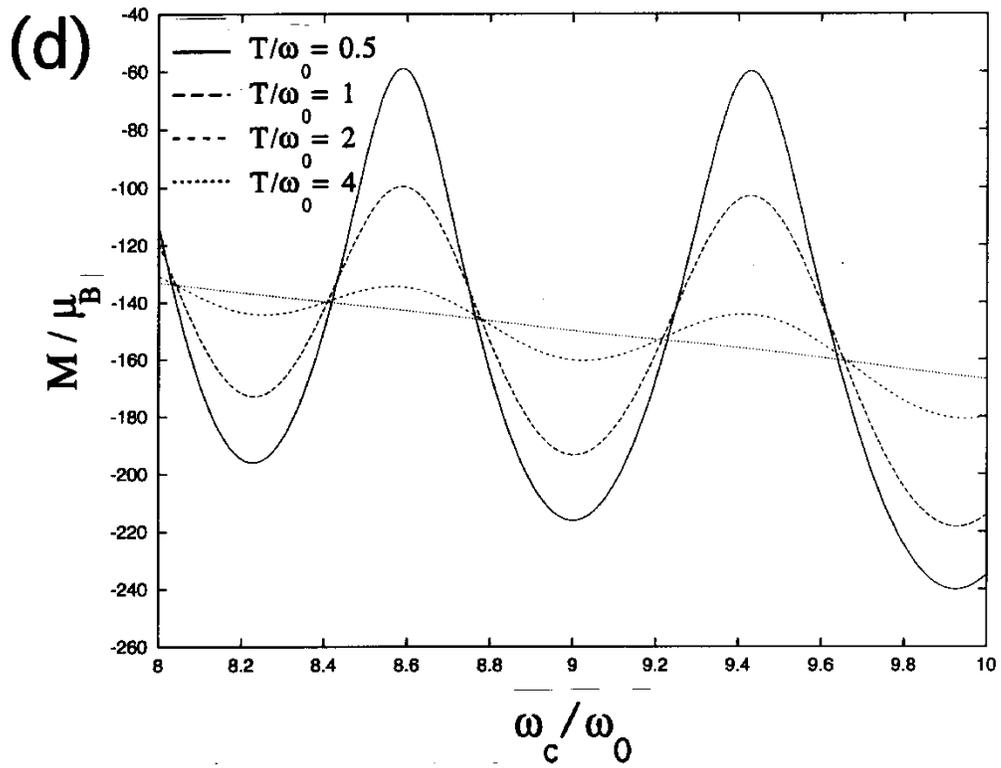
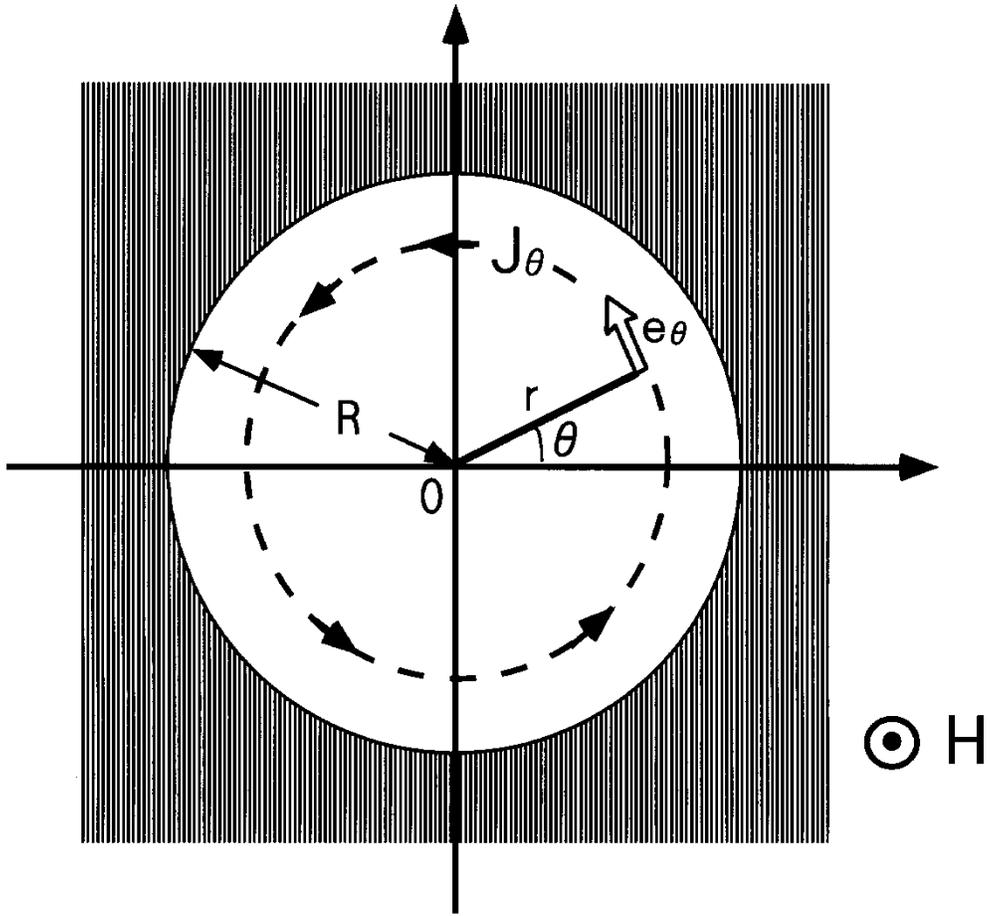
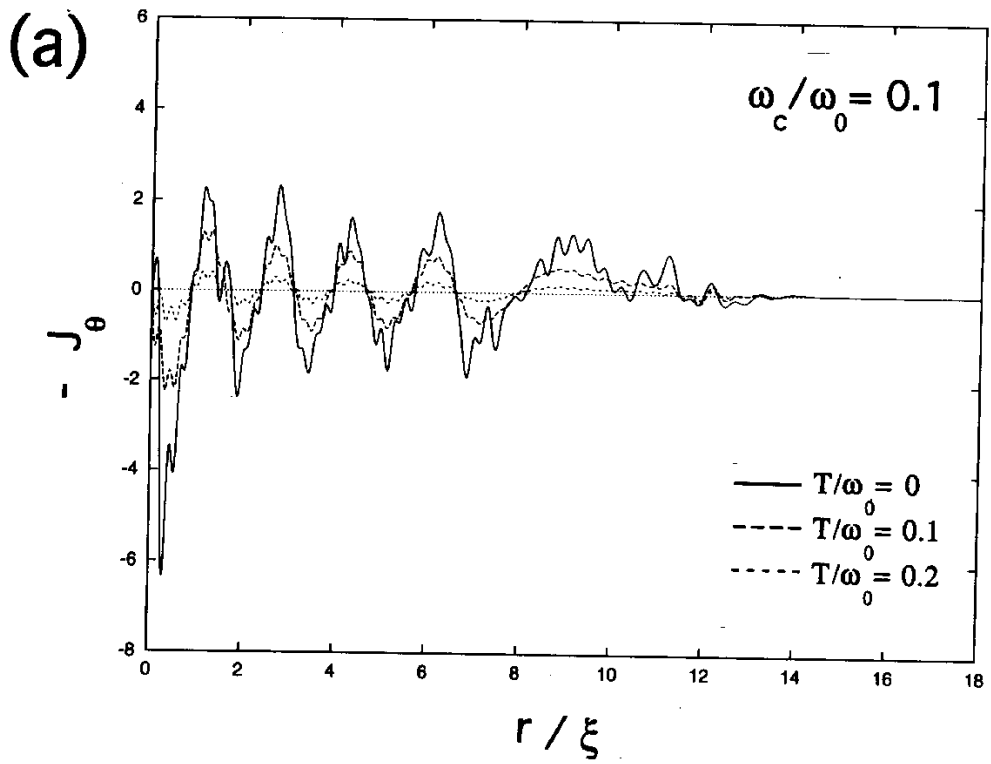
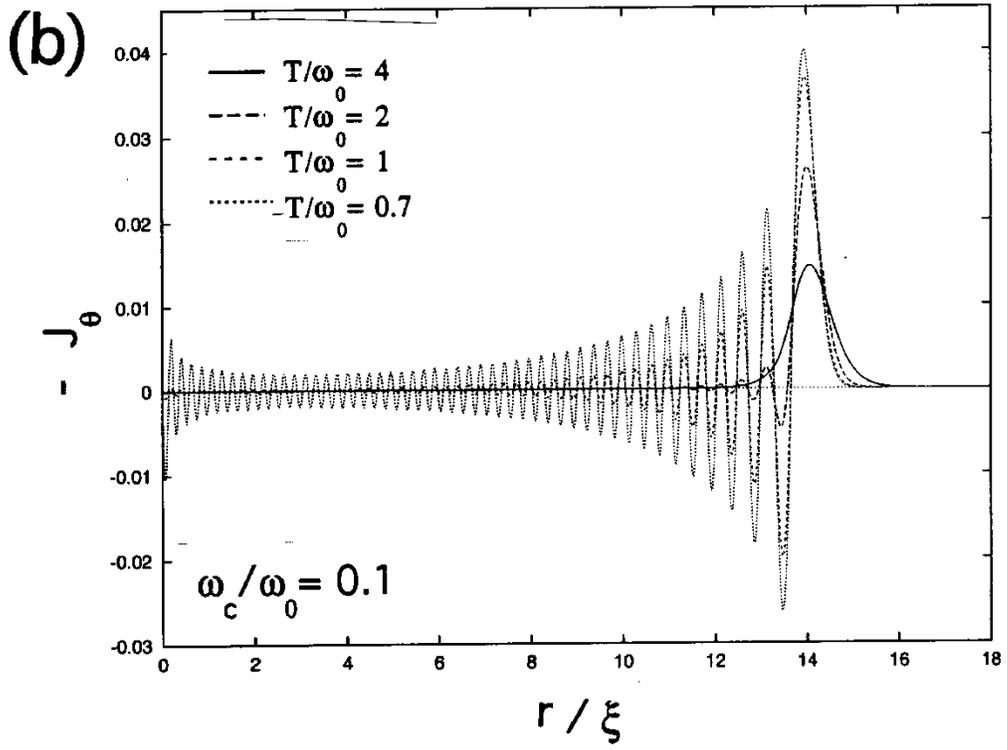
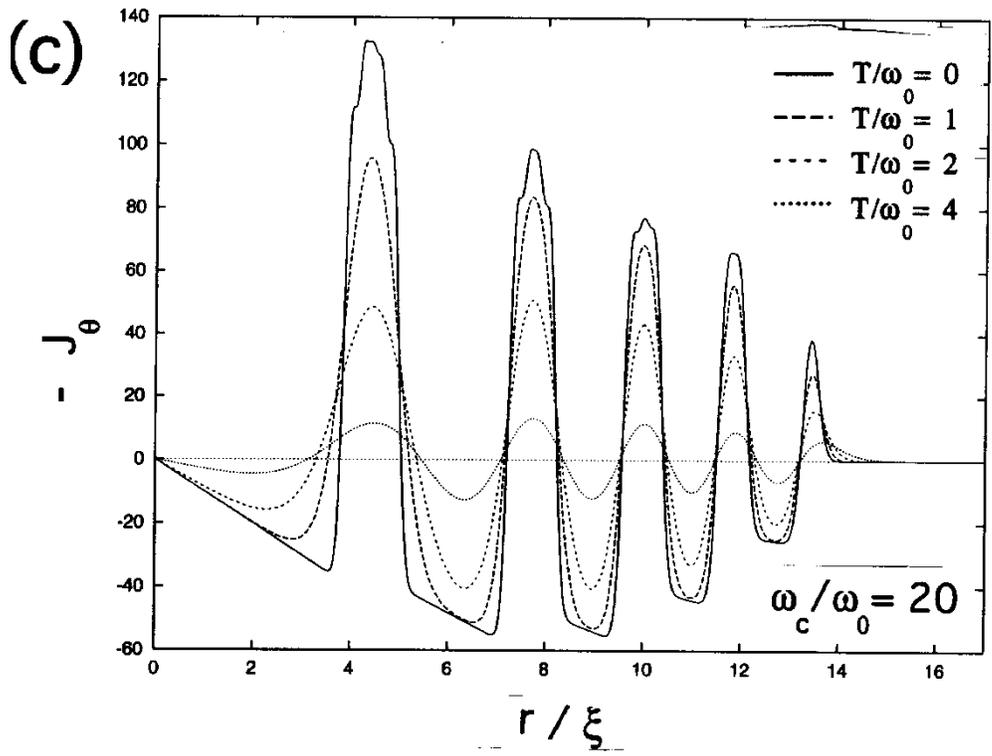
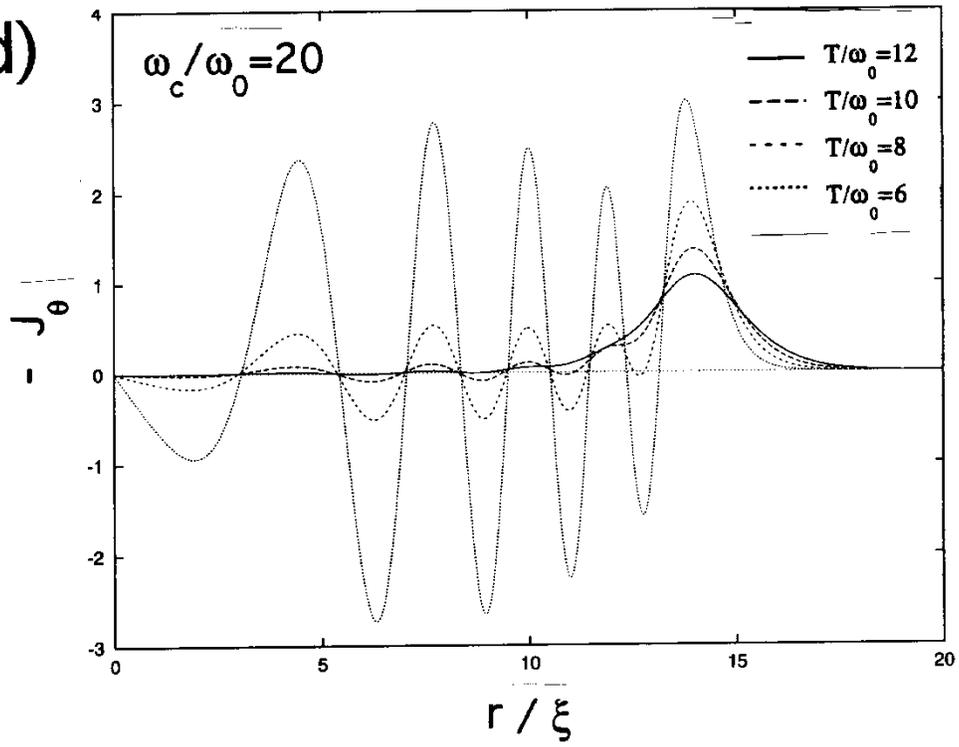
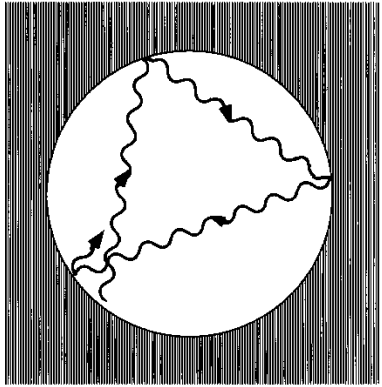
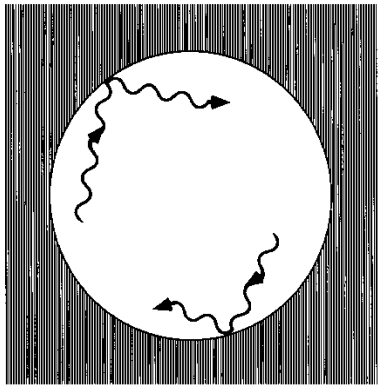
\end{document}